\documentclass[preprint,12pt]{elsarticle}

\usepackage{amssymb}
\usepackage{amsmath}
\usepackage{bm}
\usepackage{multirow}
\usepackage{threeparttable}
\usepackage{tablefootnote}
\usepackage{xcolor}
\usepackage[section]{placeins}

\usepackage{listings}
\usepackage{xcolor}
\usepackage{braket}
\definecolor{lightpink}{RGB}{255, 230, 235}

\lstdefinestyle{pythonstyle}{
    backgroundcolor=\color{lightpink},
    language=Python,
    basicstyle=\ttfamily\scriptsize, 
    keywordstyle=\color{blue},
    stringstyle=\color{orange},
    commentstyle=\color{gray},
    numbers=left,
    numberstyle=\tiny\color{gray},
    stepnumber=1,
    numbersep=5pt,
    showspaces=false,
    showstringspaces=false,
    breaklines=true,
    breakatwhitespace=true,
    tabsize=4,
    frame=single
}

\journal{Advances in Quantum Chemistry}

\begin{document}

\begin{frontmatter}

\title{Extension of the Jordan-Wigner mapping to nonorthogonal spin orbitals for quantum computing application to valence bond approaches}


\author[Eni] {Alessia Marruzzo}
\author[Poli]{Mosè Casalegno}
\author[Poli]{Piero Macchi}
\author[Eni] {Fabio Mascherpa}
\author[Eni] {Bernardino Tirri}
\author[Poli]{Guido Raos}
\author[Poli]{Alessandro Genoni\corref{cor}}
\ead{alessandro.genoni@polimi.it}
\address[Poli]{Department of Chemistry, Materials and  Chemical Engineering "Giulio Natta", Politecnico di Milano, Via Mancinelli 7, 20131 Milano, Italy} 
\address[Eni] {HPCOX/A – Advanced Modeling and Simulation, Eni S.p.A., DIT – Digital  \& Information Technology, Via Emilia 1, 20097 San Donato Milanese (MI), Italy}
\cortext[cor]{Corresponding author}
\begin{abstract}
Quantum computing offers a promising platform to address the computational challenges inherent in quantum chemistry, and particularly in valence bond (VB) methods, which are chemically appealing but suffer from high computational cost due to the use of nonorthogonal orbitals. While various fermionic-to-spin mappings exist for orthonormal spin orbitals, such as the widely used Jordan–Wigner transformations, an analogous framework for nonorthogonal spin orbitals remains undeveloped. In this work, we propose an alternative Jordan–Wigner-type mapping tailored for the nonorthogonal case, with the goal of enabling efficient quantum simulations of VB-type wavefunctions. Our approach paves the way towards the development of chemically interpretable and computationally feasible valence bond algorithms on near-term quantum devices. An initial theoretical analysis and a preliminary application demonstrate the feasibility of this encoding and its potential for extending the applicability of VB methods to larger and more complex systems.
\end{abstract}


\end{frontmatter}



\section{Introduction}\label{Intro} \footnote{This work has been accepted for publication in \emph{Advances in Quantum Chemistry} on July 12, 2025 (https://doi.org/10.1016/bs.aiq.2025.07.007). This manuscript is the author-accepted version.} 
One of the main goals of quantum chemistry is to accurately describe the electronic structure of molecules by solving the many-body Schrödinger equation \cite{Szabo1996, Helgaker2000}. To this end, a variety of approximate methods have been developed and refined over the years, making them increasingly accessible and useful even to non-specialists, often serving as a valuable complement to experimental research \cite{Levine2009, Jensen2017}. However, it is well known that traditional computational and theoretical chemistry techniques involve a trade-off between accuracy and computational feasibility. For example, Density Functional Theory (DFT) \cite{Parr1989, Becke1993, Burke2013DFT} is known to provide reliable results at a significantly lower computational cost, but its accuracy can vary depending on the choice of the exchange-correlation functional. Nevertheless, the treatment of labile intermediates and transition states, i.e. of chemical reactions, is especially challenging due to near-degeneracies (low-lying electronic states) that complicate the treatment of electron correlation effects. To also overcome this drawback, multi-determinant approaches (such as Configuration Interaction (CI), Complete Active Space Self-Consistent Field (CASSCF), Many-Body Perturbation Theory (MBPT), and Coupled Cluster (CC) methods) were developed and are recognized for their higher accuracy and reliability \cite{Helgaker2000, Shavitt2009, Olsen2011, Bartlett2007}, although their application to large systems remains computationally demanding and technically challenging.

To reduce the computational cost of the quantum chemistry techniques, different strategies have been devised. Among them, we can mention for example the multiscale embedding methods \cite{emb1} (e.g., quantum mechanics/molecular mechanics, QM/MM \cite{Warshel1976, Field1990, Senn2009}), where the most important part for the system under investigation is treated at a high quantum mechanical level of theory (for example, through a highly correlated quantum mechanical technique), while the surrounding environment is described using a less computationally intensive strategy (for instance, through molecular mechanics). This hierarchical approach enables researchers to capture essential electronic effects in the region of interest without incurring the prohibitive cost of a full quantum mechanical treatment for the whole system. Another promising avenue involves the development of linear scaling algorithms and localized orbital approaches, which take advantage of the sparsity of electronic interactions in large systems to achieve significant computational savings \cite{Goedecker1999}. More recently, machine learning (ML) and artificial intelligence (AI) techniques have been increasingly integrated into quantum chemical workflows. These approaches can be used to predict molecular properties, accelerate evaluations of the potential energy surface, and assist in the selection of functionals or basis sets, thus enhancing the traditional impact of quantum chemistry methods \cite{Rupp2012, vonLilienfeld2020}.

In this scenario, quantum computing represents another fundamentally different computational paradigm poised to transform electronic structure calculations by leveraging quantum bits (qubits) to efficiently simulate quantum systems \cite{McArdle2020, Cao2019}. Notable quantum algorithms in this area include the Quantum Phase Estimation (QPE) \cite{Abrams1997, Abrams1999, AspuruGuzik2005}, which can determine Hamiltonian eigenvalues with high precision provided that a fault-tolerant quantum hardware is available, and the Variational Quantum Eigensolver (VQE) \cite{Peruzzo2014, Tilly2022}, a hybrid quantum-classical method tailored for approximating ground-state energies on current noisy intermediate-scale quantum (NISQ) devices. Due to the current lack of fault-tolerant quantum computers, VQE remains the primary algorithm of practical use today. 

It is also worth noting that the performance of the aforementioned quantum algorithms strongly depends on the choice of the \emph{ansatz} and the design of the quantum circuit. For example, the unitary coupled-cluster (UCC) \emph{ansatz}, which is inspired by classical quantum chemistry, has been widely studied but often requires circuit depths beyond current NISQ capabilities \cite{Romero2018}. More hardware-efficient and adaptive \emph{ansätze}, such as ADAPT-VQE \cite{Grimsley2019}, have been introduced to reduce circuit complexity while maintaining chemical accuracy. In parallel, techniques like qubit tapering \cite{Bravyi2017}, active space selection \cite{Temme2017}, and error mitigation \cite{Tilly2022} are being actively developed to improve the feasibility of quantum simulations on near-term devices.

A key step in applying these algorithms to quantum chemistry problems is the mapping onto qubit operators of the electronic structure Hamiltonian, which is typically expressed in second quantization using fermionic creation and annihilation operators. This is also known as fermionic-to-spin mapping/encoding and is generally accomplished through different types of transformations, such as the Jordan–Wigner \cite{Jordan1928}, Bravyi–Kitaev \cite{Bravyi2002}, and parity mappings \cite{Seeley2012}, each of them characterized by its own advantages and disadvantages in terms of operator locality, circuit depth, and qubit efficiency. These mappings translate the fermionic Hamiltonian into a weighted sum of Pauli strings that can be measured or evolved on quantum hardware.

Exploiting the same mathematical devices, quantum computing may also expand the applicability of another important class of quantum chemistry methods, namely the valence bond (VB) approaches \cite{Shaik2011}. In fact, although VB methods offer a more intuitive connection to the traditional concepts of chemical bonding (e.g., resonance structures and their weights), they are generally associated with higher computational workloads, mainly due to the nonorthogonality of the orbitals used in the computations. This has limited the widespread adoption of VB techniques in favor of more computationally efficient molecular orbital--based methods. Therefore, a relevant goal in this field is to harness the capabilities of quantum computing to mitigate the computational demands of VB strategies, thereby enabling their application to larger and more complex systems than those for which they are currently used. Modern spin-coupled generalized valence bond (SC-GVB) approaches \cite{Cooper1991, Gerratt1997, Dunning2021} include orbital optimization and lead to very compact wavefunctions, especially compared to CASSCF ones of comparable accuracy. This compactness could be an important property within a quantum computing framework.

Anyway, two main drawbacks actually hampers the accomplishment of
the above-mentioned goal: i) the need of devising a new fermionic-to-spin
mapping for nonorthogonal spin orbitals, which is related to the use of
nonorthogonal one-electron functions (namely, orbitals) in the VB techniques;
ii) the adoption of a suitable \emph{ansatz} that could be exploited in algorithms for current NISQ devices and that keeps the inherent chemical interpretability of the valence bond philosophy. This work exclusively deals with the first aspect introducing an alternative Jordan-Wigner mapping for nonorthogonal spin orbitals as a starting point to build valence bond-like quantum computing algorithms.

The structure of the paper is as follows. Section \ref{Theory} begins with a brief overview of second quantization and of the traditional Jordan-Wigner transformation in the orthogonal case, followed by the introduction and critical examination of a novel fermionic-to-spin mapping designed for nonorthogonal spin orbitals. In Section \ref{Results}, the practical implementation of this new encoding within a quantum circuit is explored, along with its early application to a valence bond-like problem. Finally, Section \ref{Conclusions} presents concluding remarks and outlines potential directions for future research.

\section{Theory} \label{Theory}
\subsection{Jordan-Wigner mapping in the orthogonal case} \label{orto}
The application of quantum computing to quantum chemistry relies on the framework of second quantization, which ensures the antisymmetric nature of the electronic wavefunction by employing appropriately defined operators. 
These operators, which are generally referred to as creation ($\hat{a}_{p}^{+}$) and annihilation ($\hat{a}_{p}$) operators, can be used to construct any electronic state and the electronic Hamiltonian. Moreover, they must satisfy specific properties.

Given a set of spin orbitals $\left\{\phi_{i}(\mathbf{x})\right\}_{i=1}^M$, which may be orthogonal or nonorthogonal, there exists a one-to-one correspondence between the Slater determinants formed from this basis and the occupation number vector (ON-vector), which can be expressed as follows:
\begin{equation}
|\mathbf{k}\rangle=\left|k_{1} k_{2} \ldots k_{M}\right\rangle \; \; \; \text{with} \; \; k_{p}=
\left\{\begin{array}{ll}
1 & \text{if} \; \; \phi_{p} \; \; \text{occupied } \\
0 & \text{if} \; \; \phi_{p} \; \; \text{unoccupied} 
\end{array}\right.
\label{eq1}
\end{equation}
If the basis $\left\{\phi_{i}(\mathbf{x})\right\}_{i=1}^M$ is made up of orthogonal spin orbitals, the creation and annihilation operators can be defined by the following relations, respectively:
\begin{equation}
\hat{a}_{p}^{+}|\mathbf{k}\rangle=\left(1-k_{p}\right) \,\Gamma_{p}^{\mathbf{k}}\,\left|k_{1} \ldots 1_{p} \ldots k_{M}\right\rangle
    \label{eq2}
\end{equation}
and
\begin{equation}
    \hat{a}_{p}|\mathbf{k}\rangle=k_{p} \,\Gamma_{p}^{\mathbf{k}}\,\left|k_{1} \ldots 0_{p} \ldots k_{M}\right\rangle,
    \label{eq3}
\end{equation}
where, in both cases, $\Gamma_{p}^{\mathbf{k}}$ is the so-called phase factor:
\begin{equation}
\Gamma_{p}^{\mathbf{k}}=\prod_{q=1}^{p-1}(-1)^{k_{q}}
\label{eq4}
\end{equation}
For the sake of completeness, it is worth recalling that these operators allow any Slater determinant to be rewritten as follows:
\begin{equation}
 |\Psi\rangle=\prod_{i=1}^N \, \hat{a}_{i}^{+}|vac\rangle\,=\,\hat{a}_{1}^{+}\,\hat{a}_{2}^{+}\,\dots \hat{a}_{N}^{+}\, |vac\rangle
 \label{eq5}
\end{equation}
with $|vac\rangle$ as the vacuum state that disappears upon the action of any annihilation operator:
\begin{equation}
 \hat{a}_j\,|vac\rangle=0   
 \label{eq6}
\end{equation}

In the orthogonal case, the creation and annihilation operators fulfill the following anticommutation relations: 
\begin{equation}
    \left[\hat{a}_{p}^{+}, \hat{a}_{q}^{+}\right]_{+}=\hat{a}_{p}^{+}\,\hat{a}_{q}^{+}+\hat{a}_{q}^{+}\,\hat{a}_{p}^{+}=0
    \label {eq7}
\end{equation}
\begin{equation}
    \left[\hat{a}_{p}, \hat{a}_{q}\right]_{+}=\hat{a}_{p}\,\hat{a}_{q}+\hat{a}_{q}\,\hat{a}_{p}=0
    \label {eq8}
\end{equation}
\begin{equation}
    \left[\hat{a}_{p}^{+}, \hat{a}_{q}\right]_{+}=\hat{a}_{p}^{+}\,\hat{a}_{q}+\hat{a}_{q}\,\hat{a}_{p}^{+}=\delta_{pq}
    \label {eq9}
\end{equation}
leading to a simple expression of the electronic Hamiltonian:
\begin{equation}
    \hat{H}=\sum_{p, q=1}^{M} h_{p q} \;\hat{a}_{p}^{+} \hat{a}_{q}+\frac{1}{2} \sum_{p, q, r, s=1}^{M} g_{p q r s} \;\hat{a}_{p}^{+} \hat{a}_{q}^{+} \hat{a}_{s} \hat{a}_{r}
    \label{eq10}
\end{equation}
where the terms $h_{pq}$ and $g_{pqrs}$ are the one- and two-electron integrals, respectively, both defined with respect to the same orthonormal basis $\left\{\phi_{i}(\mathbf{x})\right\}_{i=1}^{M}$. Specifically, $h_{pq}$ and $g_{pqrs}$ are given by:
\begin{equation}
h_{p q}=\int d\mathbf{x} \; \phi_{p}^{*}(\mathbf{x})\left[-\frac{1}{2} \nabla^{2}-\sum_{A=1}^{N_{n u c l e i}} \frac{Z_{A}}{\left|\mathbf{r}_{i}-\mathbf{R}_{A}\right|}\right] \phi_{q}(\mathbf{x})
\label{eq11}
\end{equation}
\begin{equation}
    g_{p q r s}=\int d\mathbf{x}_{1} \, d\mathbf{x}_{2} \;\phi_{p}^{*}\left(\mathbf{x}_{1}\right) \phi_{q}^{*}\left(\mathbf{x}_{2}\right) \frac{1}{\left|\mathbf{r}_{1}-\mathbf{r}_{2}\right|} \phi_{r}\left(\mathbf{x}_{1}\right) \phi_{s}\left(\mathbf{x}_{2}\right)
    \label{eq12}
\end{equation}

In order to represent any electronic wavefunction through a qubit array for a quantum computation, a fermionic-to-spin encoding must be employed, which maps the occupation numbers of spin orbitals onto qubit states. The simplest of such encodings is the Jordan-Wigner mapping, which, in the orthogonal case, is based on the following relations:
\begin{equation}
    \hat{a}_{p}^{+} \longleftrightarrow Z_{1} \otimes Z_{2} \otimes \cdots \otimes Z_{p-1} \otimes \frac{X_{p}-i Y_{p}}{2} \otimes I_{p+1} \otimes \cdots \otimes I_{M}
    \label{eq13}
\end{equation}
and
\begin{equation}
    \hat{a}_{p}\longleftrightarrow Z_{1} \otimes Z_{2} \otimes \cdots \otimes Z_{p-1} \otimes \frac{X_{p}+i Y_{p}}{2} \otimes I_{p+1} \otimes \cdots \otimes I_{M},
    \label{eq14}
\end{equation}
where $X_{j}$, $Y_{j}$, and $Z_{j}$ are Pauli matrices/gates acting on the $j$-th qubit, while $I_{j}$ is the identity matrix (always acting on the $j$-th qubit). It is worth noting that the string of $Z$ Pauli gates $Z_{1} \otimes Z_{2} \otimes \cdots \otimes Z_{p-1}$ is included in Eqs.~\eqref{eq13} and \eqref{eq14} to account for the phase factor in Eqs.~\eqref{eq2} and \eqref{eq3}.

For completeness, one should also note that if Eqs.~\eqref{eq13} and \eqref{eq14} are used to encode the electronic Hamiltonian defined in Eq.~\eqref{eq10}, the resulting number of Pauli strings is $M^2$ for the one-electron term and $M^4$ for the two-electron term.

\subsection{Jordan-Wigner mapping in the nonorthogonal case} \label{nonorto}
To introduce the new Jordan–Wigner mapping in the nonorthogonal case, it is first of all necessary to rely on the second-quantization formalism for nonorthogonal spin orbitals, a framework that has been already extensively developed by several authors and will also be briefly reviewed below \cite{Helgaker2000,Surjan,Olsen2015,Wu1,Wu2,Wu3,Wu4}.   

Let us consider a set of nonorthogonal spin orbitals $\left\{\phi_{i}(\mathbf{x})\right\}_{i=1}^M$, with overlaps given by 
\begin{equation}
    S_{p q}=\int d\mathbf{x} \; \phi_{p}^{*}(\mathbf{x}) \;\phi_{q}(\mathbf{x}),
    \label{eq15}
\end{equation}
and let us intoduce an auxiliary set of symmetrically orthogonalized spin orbitals $\Bigl\{\tilde{\phi}_i(\mathbf{x})\Bigr\}_{i=1}^M$ obtained as
\begin{equation}
    \tilde{\phi}_{p}(\mathbf{x})=\sum_{q=1}^{M} \phi_{q}(\mathbf{x})\left[\mathbf{S}^{-\frac{1}{2}}\right]_{q p}.
    \label{eq16}
\end{equation}

Starting from the orthonormal spin orbitals defined in Eq.~\eqref{eq16}, we construct two corresponding sets of operators: the creation operators ${\hat{\tilde{a}}^+_p}$ and the annihilation operators ${\hat{\tilde{a}}_p}$. The relationship between the new creation operators ${\hat{\tilde{a}}^+_p}$ and the creation operators ${\hat{a}^+_p}$, which are associated with the initial nonorthogonal spin orbitals, mirrors the transformation between the corresponding sets of spin orbitals. Therefore, we have:
\begin{equation}
\hat{\tilde{a}}^+_p=\sum_{q=1}^M\,\hat{a}_q^+\;\bigl[\mathbf{S}^{-\frac{1}{2}}\bigr]_{qp}
\label{eq17}    
\end{equation}
and consequently
\begin{equation}
\hat{a}^+_p=\sum_{q=1}^M\,\hat{\tilde{a}}_q^+\;\bigl[\mathbf{S}^{\frac{1}{2}}\bigr]_{qp}
\label{eq18}    
\end{equation}
Now, by taking the adjoints of Eqs.~\eqref{eq17} and \eqref{eq18}, we also obtain:
\begin{equation}
(\hat{\tilde{a}}^+_p)^{\dagger}=\hat{\tilde{a}}_p=\sum_{q=1}^M\,\bigl[\mathbf{S}^{-\frac{1}{2}}\bigr]_{pq}\;\hat{a}_q
\label{eq19}    
\end{equation}
and 
\begin{equation}
(\hat{a}^+_p)^{\dagger}=\hat{a}_p=\sum_{q=1}^M\,\bigl[\mathbf{S}^{\frac{1}{2}}\bigr]_{pq}\;\hat{\tilde{a}}_q
\label{eq20}    
\end{equation}
It is important to note that, unlike $\hat{\tilde{a}}_p$, $\hat{a}_p$ given by Eq.~\eqref{eq20} is not an annihilation operator, but simply the adjoint of the creation operator $\hat{a}^+_p$.

By using Eqs.~\eqref{eq19} and \eqref{eq20} along with the anticommutation relations for the orthonormal case given by Eqs.~\eqref{eq7}, ~\eqref{eq8} and ~\eqref{eq9}, we derive the anticommutation relations for the set of operators $\{\hat{a}^+_p\}$ and $\{\hat{a}_p\}$ in the nonorthogonal regime. Notably, the first two relations remain identical to those presented in Eqs.~\eqref{eq7} and \eqref{eq8}, while the third one becomes:
\begin{equation}
\bigl[\hat{a}_p^{+} , \hat{a}_q \bigr]_{+} = \;\hat{a}_p^{+} \, \hat{a}_q + \hat{a}_q \hat{a}_p^{+}=S_{pq},
\label{eq21}
\end{equation}
with $S_{pq}$ defined through Eq.~\eqref{eq15}.

The derived anticommutation relations allow us to determine how the operators $\hat{a}_p^{+}$ and $\hat{a}_p$ act on a single ON-vector. In particular, $\hat{a}_p^{+}$ retains the same action as described by Eq.~\eqref{eq2} for the orthogonal case, whereas the action of $\hat{a}_p$ is given by:
\begin{equation}
\hat{a}_p\,|\mathbf{k}\rangle=\sum_{q=1}^{M}\,k_q\;\Gamma_q^{\mathbf{k}}\;S_{pq}\;|k_1\,\dots\,0_q\,\dots\,k_M\rangle.
\label{eq22}
\end{equation}
This expression makes it clearer that $\hat{a}_p$ is not strictly an annihilation operator in the usual sense, but rather the  adjoint of $\hat{a}_p^+$. Specifically, Eq.~\eqref{eq22} shows that applying the adjoint of a creation operator for nonorthogonal spin orbitals to an ON-vector yields a linear combination of ON-vectors, each corresponding to the removal of one electron from the original configuration. In other words, the action of $\hat{a}_p$ on an ON-vector effectively results in a weighted sum of the effects produced by multiple annihilation operators, with weights given by the overlap integrals between the nonorthogonal spin orbitals.

Building on Eqs.~\eqref{eq2} and \eqref{eq22}, we introduce a new fermionic-to-spin mapping that enables a qubit array-representation of any electronic wavefunction constructed from nonorthogonal spin orbitals. This approach can be interpreted as a generalized version of the Jordan-Wigner encoding tailored for the nonorthogonal case. Notably, the encoding for the creation operator $\hat{a}_p^+$ remains the same as in the orthonormal formulation (see Eq.~\eqref{eq13}), while the encoding for the operator $\hat{a}_p$ is given by:
\begin{equation}
\hat{a}_p \longleftrightarrow\sum_{q=1}^M\, S_{pq} \, \Bigl[Z_1 \otimes Z_2 \otimes \dots \otimes Z_{q-1} \otimes \frac{X_q+iY_q}{2} \otimes I_{q+1} \otimes \dots \otimes I_M\Bigr],
\label{eq23}
\end{equation}
with $X_{j}$, $Y_{j}$, $Z_{j}$, and $I_{j}$ having the same meaning as for Eqs.~\eqref{eq13} and \eqref{eq14}. It can be readily verified that the proposed encoding satisfies the anticommutation relations in the nonorthogonal case.

In the basis of nonorthogonal spin orbitals, the electronic Hamiltonian can be expressed as follows:
\begin{eqnarray}
 \hat{H} & = & \sum_{p,q=1}^M \; \Biggl[\,\sum_{r,s=1}^M \; \bigl[\mathbf{S}^{-1}\bigr]_{pr} \; h_{rs} \; \bigl[\mathbf{S}^{-1}\bigr]_{sq} \Biggr] \; \hat{a}_p^{+}\;\hat{a}_q  \nonumber \\
  & + & \frac{1}{2}\sum_{p,q,r,s=1}^{M}\Biggl[\,\sum_{i,j,k,l=1}^M  \, \Bigl[\mathbf{S}^{-1}\Bigr]_{pi} \, \Bigl[\mathbf{S}^{-1}\Bigr]_{qj}    \, g_{ijkl} \, \Bigl[\mathbf{S}^{-1}\Bigr]_{kr} \, \Bigl[\mathbf{S}^{-1}\Bigr]_{ls} \Biggr] \; \hat{a}_p^{+} \, \hat{a}_q^{+} \,\hat{a}_s \, \hat{a}_r \nonumber \\
  & & \label{eq24}
\end{eqnarray}
with $h_{rs}$ and $g_{ijkl}$ having the same meaning as the terms expressed via Eqs.~\eqref{eq11} and \eqref{eq12}, respectively, but in this case  using the nonorthogonal spin orbitals $\Bigl\{\phi_{i}(\mathbf{x})\Bigr\}_{i=1}^M$.

By analyzing Eq.~\eqref{eq24}, we can assess the impact of applying the new Jordan-Wigner mapping in the encoding of the electronic Hamiltonian. Specifically, it can be shown that this mapping yields up to $M^3$ Pauli strings for the one-electron term (i.e., the first term on the right-hand side of Eq.~\eqref{eq24}) and up to $M^6$ Pauli strings for the two-electron term (i.e., the second term on the right-hand side of Eq.~\eqref{eq24}). Compared to the orthogonal case (see Subsection \ref{orto}), this results in a significantly higher number of Pauli strings, which may negatively affect the performance of quantum simulations.

To address this drawback, one can employ the so-called biorthogonal spin orbitals $\Bigl\{\overline{\phi}_{i}(\mathbf{x})\Bigr\}_{i=1}^M$, which are related to the original nonorthogonal spin orbitals via the following transformation:
\begin{equation}
  \overline{\phi}_p(\mathbf{x})=\sum_{q=1}^M\,\phi_q(\mathbf{x})\,\Bigl[\mathbf{S}^{-1}\Bigr]_{qp}
  \label{eq25}
\end{equation}
Furthermore, from Eq.~\eqref{eq25}, the corresponding biorthogonal  operators $\left\{\hat{\overline{a}}_p\right\}$ are naturally defined as follows:
\begin{equation}
 \hat{\overline{a}}_p=\sum_{q=1}^M\,\Bigl[\mathbf{S}^{-1}\Bigr]_{pq} \, \hat{a}_q
 \label{eq26}
\end{equation} 
It is worth noting that, although the biorthogonal operators $\left\{\hat{\overline{a}}_p\right\}$  are not the adjoints of the corresponding creation operators, they serve as the proper annihilation operators associated with the nonorthogonal spin orbitals. Moreover, they fulfill the following anticommutation relation:
\begin{equation}
 \bigl[\hat{a}_p^{+} , \hat{\overline{a}}_q \bigr]_{+} = \;\hat{a}_p^{+} \, \hat{\overline{a}}_q + \hat{\overline{a}}_q \hat{a}_p^{+}=\delta_{pq}
 \label{eq27}
\end{equation}

By leveraging the definitions of biorthogonal spin orbitals and of the corresponding operators (see Eqs.~\eqref{eq25} and \eqref{eq26}, respectively), the electronic Hamiltonian given in Eq.~\eqref{eq24} can be reformulated as follows:
\begin{equation}
  \hat{H}=\sum_{p,q=1}^M\,h_{\,\overline{p}\,q} \, \hat{a}_p^{+} \, \hat{\overline{a}}_q + \frac{1}{2}\sum_{p,q,r,s=1}^M \, g_{\,\overline{p}\,\overline{q}\,r\,s} \, \hat{a}_p^{+} \, \hat{a}_q^{+} \, \hat{\overline{a}}_s \, \hat{\overline{a}}_r
  \label{eq28}
\end{equation}
where the one- and two-electron integrals now become:
\begin{equation}
h_{\,\overline{p}\,q}=\int d\mathbf{x} \; \overline{\phi}^{\;*}_p(\mathbf{x})\left[-\frac{1}{2} \nabla^{2}-\sum_{A=1}^{N_{n u c l e i}} \frac{Z_{A}}{\left|\mathbf{r}_{i}-\mathbf{R}_{A}\right|}\right] \phi_{q}(\mathbf{x}) 
\label{eq29}
\end{equation}
and
\begin{equation}
g_{\,\overline{p}\,\overline{q}\,r\,s}\int d\mathbf{x}_{1} \, d\mathbf{x}_{2} \;\overline{\phi}_{p}^{\;*}\left(\mathbf{x}_{1}\right) \overline{\phi}_{q}^{\;*}\left(\mathbf{x}_{2}\right) \;\frac{1}{\left|\mathbf{r}_{1}-\mathbf{r}_{2}\right|} \;\phi_{r}\left(\mathbf{x}_{1}\right) \phi_{s}\left(\mathbf{x}_{2}\right)
    \label{eq30}
\end{equation}
with $h_{\,\overline{p}\,q} \neq h_{\,\overline{q}\,p}$ and $g_{\,\overline{p}\,\overline{q}\,r\,s} \neq g_{\,\overline{r}\,\overline{s}\,p\,q}$, which results in a larger number of integrals to be computed and stored.

Based on the definition provided in Eq.~\eqref{eq26} and the proposed Jordan-Wigner mapping for the adjoint operators $\left\{\hat{a}_p\right\}$ (see Eq.~\eqref{eq23}), we can extend the approach to the biorthogonal operators by introducing the following Jordan-Wigner-like encoding:
\begin{equation}
 \hat{\overline{a}}_{p}\longleftrightarrow Z_{1} \otimes Z_{2} \otimes \cdots \otimes Z_{p-1} \otimes \frac{X_{p}+i Y_{p}}{2} \otimes I_{p+1} \otimes \cdots \otimes I_{M},
 \label{eq31} 
\end{equation}
which is analogous to the mapping for the annihilation operator in the orthogonal case.

When the mapping defined in Eq.~\eqref{eq31} is applied to the electronic Hamiltonian in Eq.~\eqref{eq28}, the number of resulting Pauli strings is reduced to $M^2$ for the one-electron term and $M^4$ for the two-electron term. This matches the scaling observed with the conventional Jordan-Wigner mapping applied to orthogonal spin orbitals. However, in this case, there is a slightly higher overhead due to the increased number of one- and two-electron integrals that must be computed and stored (see the inequalities below Eq.~\eqref{eq30}).

In conclusion, when working with wavefunctions constructed from nonorthogonal spin orbitals, it is advisable to use the Jordan-Wigner-like mappings defined in Eqs.~\eqref{eq13} and \eqref{eq23} for the encoding and manipulation of the wavefunctions. However, the encoding of the electronic Hamiltonian should be addressed by the mappings given in Eqs.~\eqref{eq13} and \eqref{eq31}.

\section{Implementation and Preliminary Results} 
\label{Results}
\subsection{Implementation in PennyLane}
In order to assess the accuracy of the proposed Jordan–Wigner encoding in the nonorthogonal case, the mappings given by Eqs.~\eqref{eq13}, \eqref{eq23}, and \eqref{eq31} were implemented numerically using the \texttt{PennyLane} library (https://pennylane.ai/), but other quantum computing frameworks could be equivalently exploited. This was done by cloning version \texttt{v0.41.1} of \texttt{PennyLane} and adding an option to account for nonorthogonal spin orbitals in the code for the Jordan–Wigner transformation function. In particular, the operators $\left\{\hat{a}_p\right\}$ were re-encoded as defined in Eq.~\eqref{eq23} whenever a nontrivial overlap matrix $S_{pq}$ is given. Furthermore, to fully exploit \texttt{PennyLane}'s built-in classes and functions, we used the highly-optimized \texttt{PennyLane-Lightning} backend to perform our calculations.

For the sake of clarity, we now provide a brief introduction to the key \texttt{PennyLane} components used in our implementation. \texttt{PennyLane} has built-in classes for Fermi operators: \texttt{FermiA} for annihilation operators and \texttt{FermiC} for creation operators. For example, \texttt{FermiC(0)} is a creation operator acting on qubit 0. Multiplication of Fermi operators gives a \texttt{FermiWord}, while a linear combination of multiple \texttt{FermiWords} is called a \texttt{FermiSentence}: for instance, the operator
$
\hat{a}_p^{+}  \hat{a}_q + \hat{a}_p^{+}  \hat{a}_q^{+} \hat{a}_s \hat{a}_r,
$
is a \texttt{FermiSentence} made up of the \texttt{FermiWords}
$\hat{a}_p^{+}  \hat{a}_q$ and $\hat{a}_p^{+}  \hat{a}_q^{+} \hat{a}_s \hat{a}_r$.

The implementation of the mapping for the operators $\left\{\hat{a}_p\right\}$ defined by Eq.~\eqref{eq23} is specifically given by the following sample \texttt{Python} code for the definition of the new function \texttt{a\_operator}:
\newpage
\begin{lstlisting}[style=pythonstyle, language=Python, caption=Sample \texttt{Python} code schematically showing the definition of the new function \texttt{a\_operator}., label=lst:samplecode1]
def a_operator(
    spin_orb: int, 
    s_pq: np.array) -> FermiSentence:
    a_spinorb = s_pq[0]*FermiA(0)
    for index_spin_orb in range(1,s_pq.shape[0]):
        a_spinorb = a_spinorb + 
                    s_pq[index_spin_orb]*FermiA(index_spin_orb)
\end{lstlisting}

\noindent where \texttt{s\_pq} represents one row of the overlap matrix defined through Eq.~\eqref{eq15}, while \texttt{a\_spinorb} corresponds to the \texttt{FermiSentence} given by Eq.~\eqref{eq23}.

The newly defined function \texttt{a\_operator} was subsequently used to modify \texttt{jordan\_wigner}, a built-in function used by \texttt{PennyLane} in order to convert \texttt{FermiWords} and \texttt{FermiSentences} into Pauli strings. Note that \texttt{jordan\_wigner} can accept either a \texttt{FermiWord} or a \texttt{FermiSentence} as input. In the latter case, the function iterates over the individual \texttt{FermiWords} making up the \texttt{FermiSentence}. Therefore, to better leverage the optimization of \texttt{PennyLane}'s code, we modified the built-in \texttt{jordan\_wigner} function only for the case in which the input is a \texttt{FermiWord}, as schematically shown by the following sample \texttt{Python} code:

\begin{lstlisting}[style=pythonstyle, language=Python, caption=Sample \texttt{Python} code schematically showing the modification of the built-in function \texttt{jordan\_wigner} in \texttt{PennyLane} when its argument is a \texttt{FermiWord}., label=lst:samplecode2]
def jordan_wigner(fermi_operator: FermiWord, 
                  matrix_overlap=None) -> PauliSentence:
    # wires is a list with the orbital indices 
    wires = list(fermi_operator.wires)
    if np.any(matrix_overlap):
        # If there is the matrix overlap 
        #    we have to deal with non-orthogonal orbitals
        fs_new=1.0
        for item in fermi_operator.items():
            # item is a tuple
            (wire,sign) = item 
            # sign is - for FermiA and + for FermiC
            if sign == "-":
                # if  FermiA I call a_operator defined previously
                new_a = a_operator(wire, matrix_overlap[wire])
                fs_new = fs_new*new_a
            else:
                fs_new = fs_new*FermiC(wire)
        # Call again the jordan_wigner mapping
        #    on the new FermiSentence fs_new
        qubit_operator = jordan_wigner(fs_new,matrix_overlap=None)
    else:
        # Usual PennyLane built-in function is called 
        #     if no overlap matrix is given
        ...
\end{lstlisting}

As shown in the sample code above, when analyzing the string of operators that make up the \texttt{FermiWord}, if an operator ${\hat{a}_p}$ is detected, a corresponding \texttt{FermiSentence} is constructed by calling the new function \texttt{a\_operator}. Conversely, when the program encounters an operator $\hat{a}_p^{+}$ or a an operator $\hat{\overline{a}}_p$ in the string of the \texttt{FermiWord}, it uses the standard built-in classes \texttt{FermiC} and \texttt{FermiA} to encode the creation and annihilation operators, respectively.

\subsection{Test calculations} \label{test}
\begin{figure}
\includegraphics[width=\textwidth]{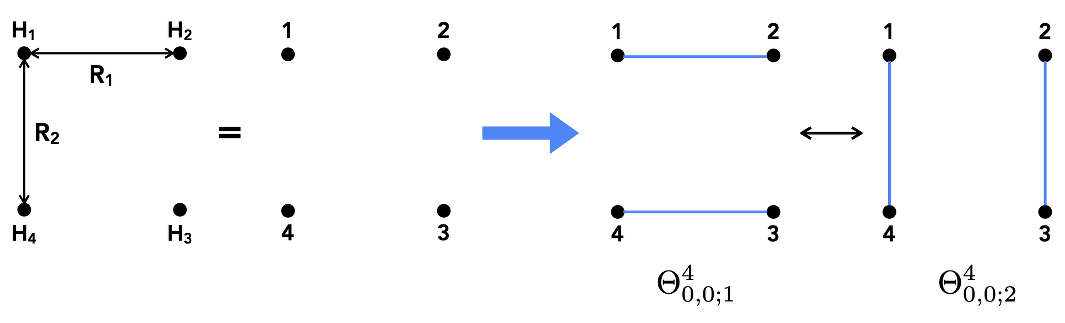}
  \caption{H$_4$ and its possible spin-coupling modes in the singlet state, with blue lines indicating the spin-couplings between two electrons. The $R_1$ and $R_2$ parameters for the definition of the adopted geometry are also indicated.}  
  \label{fig1}
\end{figure}

To evaluate the implementation of the proposed mapping, we initially considered the simple H$_4$ system, where the four hydrogen atoms were placed at the vertices of a square with a side length of 0.850 \AA (namely, $R_1=R_2=0.850$ \AA; see Fig.~\ref{fig1}). For simplicity, we employed the minimal STO-3G basis set. These basis functions were considered as fixed nonorthogonal orbitals (indicated below as $\phi_1$, $\phi_2$, $\phi_3$ and $\phi_4$) to construct the spin-coupled generalized valence bond (SCGVB) wavefunction \cite{Cooper1991, Gerratt1997, Dunning2021} for the examined system.

In particular, since we considered a system of 4 electrons in a singlet state, only two spin-coupling modes are possible (see Fig.~\ref{fig1}). Therefore, the global wavefunction can be written as a normalized linear combination of two structures (notably, $\Psi_{0,0;1}^4$ and $\Psi_{0,0;2}^4$) as follows:
\begin{equation}
  \Psi_{SCGVB} = \mathcal{N}\;\Bigl[C_1 \; \Psi_{0,0;1}^4 \,+ \, C_2 \; \Psi_{0,0;2}^4 \Bigr] \label{eq32}
\end{equation}
where $\mathcal{N}$ is the proper normalization constant, and the two structures $\psi_{0,0;1}^4$ and $\psi_{0,0;2}^4$ can be explicitly written like this:
\begin{equation}
\Psi_{0,0;1}^4 = \hat{A}\Bigl(\phi_1 \,\phi_2\, \phi_3\, \phi_4\, \Theta_{0,0;1}^4\Bigr) \label{eq33}
\end{equation}
and
\begin{equation}
 \Psi_{0,0;2}^4 =\hat{A}\Bigl(\phi_1 \,\phi_2\, \phi_3\, \phi_4\, \Theta_{0,0;2}^4\Bigr) \label{eq34}
\end{equation}

In Eqs.~\eqref{eq33} and \eqref{eq34}, $\Theta_{0,0;1}^4$ and $\Theta_{0,0;2}^4$  are the two spin eigenfunctions associated with the two above-mentioned spin-coupling modes. In particular, by adopting the Rumer basis and following the spin-coupling scheme depicted in Fig. \ref{fig1}, they are given by:
\begin{equation}
 \Theta_{0,0;1}^4=\frac{1}{\sqrt{2}}\Bigl(\alpha(\sigma_1)\,\beta(\sigma_2)-\beta(\sigma_1)\,\alpha(\sigma_2)\Bigr)\; \frac{1}{\sqrt{2}}\Bigl(\alpha(\sigma_3)\,\beta(\sigma_4)-\beta(\sigma_3)\,\alpha(\sigma_4)\Bigr)
\label{eq35}
\end{equation}
and
\begin{equation}
 \Theta_{0,0;2}^4=\frac{1}{\sqrt{2}}\Bigl(\alpha(\sigma_1)\,\beta(\sigma_4)-\beta(\sigma_1)\,\alpha(\sigma_4)\Bigr)\; \frac{1}{\sqrt{2}}\Bigl(\alpha(\sigma_2)\,\beta(\sigma_3)-\beta(\sigma_2)\,\alpha(\sigma_3)\Bigr),
 \label{eq36}
\end{equation}
with the first spin eigenfunction $\Theta_{0,0;1}^4$ corresponding to the (1,2) and (3,4) electron pairings, and the second spin eigenfunction $\Theta_{0,0;2}^4$ associated with the (1,4) and (2,3) electron couplings.

Bearing in mind that the application of the antisymmetrizer $\hat{A}$ to a string of spin orbitals produces a Slater determinant and exploiting Eqs.~\eqref{eq35} and \eqref{eq36} into Eqs.~\eqref{eq33} and \eqref{eq34}, the two spin-coupled structures can be rewritten like this:
\begin{equation}
  \Psi_{0,0;1}^4  =  \frac{1}{2\sqrt{4!}} \; \Biggl[\Big|\phi_1 \,\overline{\phi}_2\, \phi_3\, \overline{\phi}_4\Big| - \Big|\phi_1 \,\overline{\phi}_2\, \overline{\phi}_3\, \phi_4\Big|-\Big|\overline{\phi}_1 \,\phi_2\, \phi_3\, \overline{\phi}_4\Big|+\Big|\overline{\phi}_1 \,\phi_2\, \overline{\phi}_3\, \phi_4\Big|\Biggr]
   \label{eq37}
\end{equation}
and
\begin{equation}
  \Psi_{0,0;2}^4 = \frac{1}{2\sqrt{4!}} \; \Biggl[\Big|\phi_1 \,\phi_2\, \overline{\phi}_3\, \overline{\phi}_4\Big| - \Big|\phi_1 \,\overline{\phi}_2\, \phi_3\, \overline{\phi}_4\Big|-\Big|\overline{\phi}_1 \,\phi_2\, \overline{\phi}_3\, \phi_4\Big|+\Big|\overline{\phi}_1 \,\overline{\phi}_2\, \phi_3\, \phi_4\Big|\Biggr],
  \label{eq38}
\end{equation}
where $\phi_i$ indicates a spatial orbital paired with a spin function $\alpha$, while $\overline{\phi}_i$ indicates a spatial orbital paired with a spin function $\beta$.

By analyzing Eqs.~\eqref{eq37} and \eqref{eq38}, it is easy to observe that the two structures can be expressed only in terms of six unique unnormalized Slater determinants, which can be labeled as follows:
\begin{equation}
 \psi_1  =  \Big|\phi_1 \,\overline{\phi}_2\, \phi_3\, \overline{\phi}_4\Big|
\label{eq39}
\end{equation}
\begin{equation}
 \psi_2 = \Big|\phi_1 \,\overline{\phi}_2\, \overline{\phi}_3\, \phi_4\Big|
\label{eq40}
\end{equation}
\begin{equation}
 \psi_3  =  \Big|\overline{\phi}_1 \,\phi_2\, \phi_3\, \overline{\phi}_4\Big|
\label{eq41}
\end{equation}
\begin{equation}
\psi_4  =  \Big|\overline{\phi}_1 \,\phi_2\, \overline{\phi}_3\, \phi_4\Big|
\label{eq42}
\end{equation}
\begin{equation}
\psi_5  =  \Big|\phi_1 \,\phi_2\, \overline{\phi}_3\, \overline{\phi}_4\Big|
\label{eq43}
\end{equation}
\begin{equation}
\psi_6  =  \Big|\overline{\phi}_1 \,\overline{\phi}_2\, \phi_3\, \phi_4\Big|
\label{eq44}
\end{equation}
Therefore, the two structures $\Psi_{0,0;1}^4$ and $\Psi_{0,0;2}^4$ become
\begin{equation}
  \Psi_{0,0;1}^4  =  \frac{1}{2\sqrt{4!}} \; \Bigl[\psi_1-\psi_2-\psi_3+\psi_4\Bigr]
   \label{eq45}
\end{equation}
and
\begin{equation}
  \Psi_{0,0;2}^4 = \frac{1}{2\sqrt{4!}} \; \Bigl[\psi_5-\psi_1-\psi_4+\psi_6\Bigr].
  \label{eq46}
\end{equation}
The ultimate goal is to determine the coefficients $C_1$ and $C_2$ in Eq.~\eqref{eq32}. This requires solving the following generalized eigenvalue problem:
\begin{equation}
  \textbf{H}\,\textbf{C}\;=\,\textbf{P}\,\textbf{C}\,\textbf{E}, 
  \label{eq47}
\end{equation}
where $\textbf{H}$ and $\textbf{P}$ are respectively the Hamiltonian and overlap matrices expressed in the basis formed by the structures $\Psi_{0,0;1}^4$ and  $\Psi_{0,0;2}^4$. For the sake of completeness, it is also worth noting that, due to the nonorthogonality of the structures, the weights of $\Psi_{0,0;1}^4$ and  $\Psi_{0,0;2}^4$ are given by the Chirgwin-Coulson coefficients that have this general definition:
\begin{equation}
  W_{i}\,=\,C_i\,\sum_{j}C_j\,P_{ij}
  \label{eq48}
\end{equation}

As an initial step, to verify both correctness and implementation of the proposed Jordan-Wigner mapping for nonorthogonal spin orbitals, we computed the Hamiltonian matrix elements $\langle \psi_i | \hat{H} | \psi_j \rangle$ and the overlap integrals $\langle \psi_i | \psi_j\rangle$ using the basis of the six distinct Slater determinants $\bigl\{\psi_i\bigr\}_{i=1}^6$ specified above. These computations were carried out using the newly proposed Jordan-Wigner encoding and the results were subsequently compared with the outcomes obtained through conventional methods, specifically by using the standard L\"{o}wdin rules \cite{Lowdin1955QTMP, McWeeny1989MQM} for the evaluation of matrix elements between Slater determinants built from nonorthogonal spin orbitals. The results of both computational approaches are presented in Tables \ref{tab1} and \ref{tab2}. It is readily apparent that the proposed fermionic-to-spin mapping for nonorthogonal spin orbitals yields results identical to those obtained using the conventional Löwdin rules, thereby confirming its validity and demonstrating its potential for future valence bond-inspired quantum computations.

\begin{table}
 \small
  \caption{Values (in atomic units) of the Hamiltonian matrix elements between unique Slater determinants as obtained through the new fermionic-to-spin Jordan-Wigner (JW) encoding for nonorthogonal orbitals and the usual L\"{o}wdin rules. All values refer to the square geometry of H$_4$ with side length of 0.850 \AA}
\label{tab1}
\vspace{2.5mm}
\centering
 \begin{threeparttable}
 \begin{tabular}{lcc}
    \hline 
    Matrix elements   &  New JW encoding &  L\"{o}wdin rules \\  
    \hline  
$\langle \psi_1 | \hat{H} | \psi_1 \rangle$ & -3.6717179032803267 &-3.6717179032803271 \\
$\langle \psi_2 | \hat{H} | \psi_1 \rangle$ &  0.7097218084667191 & 0.7097218084667193 \\
$\langle \psi_2 | \hat{H} | \psi_2 \rangle$ & -2.0077738595167930 &-2.0077738595167935 \\
$\langle \psi_3 | \hat{H} | \psi_1 \rangle$ &  0.7097218084667178 & 0.7097218084667183 \\
$\langle \psi_3 | \hat{H} | \psi_2 \rangle$ & -0.2415263736519134 &-0.2415263736519131 \\
$\langle \psi_3 | \hat{H} | \psi_3 \rangle$ & -2.0077738595167930 &-2.0077738595167935 \\
$\langle \psi_4 | \hat{H} | \psi_1 \rangle$ &  0.0076442524542684 & 0.0076442524542684 \\
$\langle \psi_4 | \hat{H} | \psi_2 \rangle$ &  0.7097218084667181 & 0.7097218084667184 \\
$\langle \psi_4 | \hat{H} | \psi_3 \rangle$ &  0.7097218084667194 & 0.7097218084667193 \\
$\langle \psi_4 | \hat{H} | \psi_4 \rangle$ & -3.6717179032803267 &-3.6717179032803271 \\
$\langle \psi_5 | \hat{H} | \psi_1 \rangle$ &  0.7097218084667177 & 0.7097218084667183 \\
$\langle \psi_5 | \hat{H} | \psi_2 \rangle$ &  0.0023350996380428 & 0.0023350996380428 \\
$\langle \psi_5 | \hat{H} | \psi_3 \rangle$ &  0.0023350996380429 & 0.0023350996380428 \\
$\langle \psi_5 | \hat{H} | \psi_4 \rangle$ &  0.7097218084667191 & 0.7097218084667191 \\
$\langle \psi_5 | \hat{H} | \psi_5 \rangle$ & -2.0077738595167927 &-2.0077738595167940 \\
$\langle \psi_6 | \hat{H} | \psi_1 \rangle$ &  0.7097218084667191 & 0.7097218084667191 \\
$\langle \psi_6 | \hat{H} | \psi_2 \rangle$ &  0.0023350996380429 & 0.0023350996380428 \\
$\langle \psi_6 | \hat{H} | \psi_3 \rangle$ &  0.0023350996380428 & 0.0023350996380428 \\
$\langle \psi_6 | \hat{H} | \psi_4 \rangle$ &  0.7097218084667177 & 0.7097218084667183 \\
$\langle \psi_6 | \hat{H} | \psi_5 \rangle$ & -0.2415263736519133 &-0.2415263736519131 \\
$\langle \psi_6 | \hat{H} | \psi_6 \rangle$ & -2.0077738595167927 &-2.0077738595167940 \\
  \hline
\end{tabular}
\end{threeparttable}
\end{table}

\begin{table}
 \small
  \caption{Values (in atomic units) of the overlap matrix elements between unique Slater determinants as obtained through the new fermionic-to-spin Jordan-Wigner (JW) encoding for nonorthogonal orbitals and the usual L\"{o}wdin rules. All values refer to the square geometry of H$_4$ with side length of 0.850 \AA}
\label{tab2}
\vspace{2.5mm}
\centering
 \begin{threeparttable}
 \begin{tabular}{lcc}
    \hline 
    Matrix elements    &  New JW encoding &  L\"{o}wdin rules \\  
    \hline  
    $\langle \psi_1 | \psi_1 \rangle$ &  0.7242249455030644 &  0.7242249455030643    \\ 
    $\langle \psi_2 | \psi_1 \rangle$ & -0.1306613259134710 & -0.1306613259134711    \\ 
    $\langle \psi_2 | \psi_2 \rangle$ &  0.4269698246566512 &  0.4269698246566513    \\ 
    $\langle \psi_3 | \psi_1 \rangle$ & -0.1306613259134708 & -0.1306613259134709    \\ 
    $\langle \psi_3 | \psi_2 \rangle$ &  0.0390397486030691 &  0.0390397486030691    \\ 
    $\langle \psi_3 | \psi_3 \rangle$ &  0.4269698246566512 &  0.4269698246566513    \\ 
    $\langle \psi_4 | \psi_1 \rangle$ &  0.0000000000000000 &  0.0000000000000000    \\ 
    $\langle \psi_4 | \psi_2 \rangle$ & -0.1306613259134708 & -0.1306613259134709    \\ 
    $\langle \psi_4 | \psi_3 \rangle$ & -0.1306613259134710 & -0.1306613259134711    \\ 
    $\langle \psi_4 | \psi_4 \rangle$ &  0.7242249455030644 &  0.7242249455030643    \\ 
    $\langle \psi_5 | \psi_1 \rangle$ & -0.1306613259134708 & -0.1306613259134709    \\ 
    $\langle \psi_5 | \psi_2 \rangle$ & -0.0015536397917990 & -0.0015536397917990    \\ 
    $\langle \psi_5 | \psi_3 \rangle$ & -0.0015536397917990 & -0.0015536397917990    \\ 
    $\langle \psi_5 | \psi_4 \rangle$ & -0.1306613259134710 & -0.1306613259134710    \\
    $\langle \psi_5 | \psi_5 \rangle$ &  0.4269698246566512 &  0.4269698246566513    \\ 
    $\langle \psi_6 | \psi_1 \rangle$ & -0.1306613259134710 & -0.1306613259134710    \\ 
    $\langle \psi_6 | \psi_2 \rangle$ & -0.0015536397917990 & -0.0015536397917990    \\ 
    $\langle \psi_6 | \psi_3 \rangle$ & -0.0015536397917990 & -0.0015536397917990    \\
    $\langle \psi_6 | \psi_4 \rangle$ & -0.1306613259134708 & -0.1306613259134709    \\ 
    $\langle \psi_6 | \psi_5 \rangle$ &  0.0390397486030691 &  0.0390397486030691    \\
    $\langle \psi_6 | \psi_6 \rangle$ &  0.4269698246566512 &  0.4269698246566513    \\
     \hline 
  \end{tabular}
\end{threeparttable}
\end{table}

\newpage

From the matrix element values obtained through the new Jordan-Wigner mapping (see the second column of Tables \ref{tab1} and \ref{tab2}) we have afterwards computed the Hamiltonian and overlap matrix elements between the structures $\Psi_{0,0;1}^4$ and $\Psi_{0,0;2}^4$. By solving the generalized eigenvalue problem represented by Eq.~\eqref{eq47} we have thus determined the coefficients and the Chirgwin-Coulson weights corresponding to the above-mentioned structures. We obtained coefficients $C_1$ and $C_2$ equal to -2.8491 and 2.8491, respectively. The obtained Chirgwin-Coulson weights $W_1$ and $W_2$ resulted exactly equal to 0.5, which are consistent with the square-geometry adopted in these first test calculations.

The values are reported in Table \ref{tab3}, along with the results obtained for alternative geometries of the H$_4$ system under investigation. In addition to the square geometry with a side length of 0.850 Å (referred to as Square 1 in Table \ref{tab1}), we initially examined a slightly stretched rectangular configuration (Rectangle 1 in Table \ref{tab1}). In this case, the Chirgwin–Coulson weights of the two contributing structures are no longer equivalent, with structure $\Psi_{0,0;1}^4$ dominating over structure $\Psi_{0,0;2}^4$. This observation is consistent with the fact that distance $R_1$ is shorter than $R_2$ (see also Fig. \ref{fig1}). By interchanging the values of $R_1$ and $R_2$, we generated a second slightly stretched rectangular geometry (Rectangle 1B in Table \ref{tab1}). In this geometrical configuration, structure $\Psi_{0,0;2}^4$ is predominant over $\Psi_{0,0;1}^4$, with the Chirgwin–Coulson weights effectively swapped compared to the Rectangle 1 case. This reversal again reflects the modified structural parameters. Finally, by further distorting the geometry, we considered three additional rectangular configurations (Rectangles 2, 3, and 4 in Table \ref{tab1}). In these cases, we consistently observe an increasing predominance of structure $\Psi_{0,0;1}^4$ at the expense of $\Psi_{0,0;2}^4$, which correlates directly with the growing difference between distances $R_1$ and $R_2$.

\begin{table}
 \small
  \caption{Coefficients ($C_1$ and $C_2$) and Chirgwin-Coulson weights ($W_i$ and $W_2$) associated with the structures $\Psi_{0,0;1}^4$ and $\Psi_{0,0;2}^4$, as obtained for different geometries of the H$_4$ system (see distances $R_1$ and $R_2$)}
\label{tab3}
\vspace{2.5mm}
\centering
 \begin{threeparttable}
 \begin{tabular}{cccccccccc}
    \hline 
    Geometry & $R_1$ (\AA) & $R_2$ (\AA) &  & $C_1$ &  $C_2$ & &  $W_1$ & $W_2$  \\  
    \hline  
    Square 1     & 0.850 & 0.850 & & -2.8491 &  2.8491 & & 0.5000 &  0.5000   \\ 
    Rectangle 1  & 0.825 & 0.875 & & -3.6051 &  1.9966 & & 0.6781 &  0.3219   \\ 
    Rectangle 1B & 0.875 & 0.825 & &  1.9966 & -3.6051 & & 0.3219 &  0.6781   \\ 
    Rectangle 2  & 0.800 & 1.000 & & -4.3316 &  0.4927 & & 0.9325 &  0.0675   \\
    Rectangle 3  & 0.775 & 1.050 & & -4.3303 &  0.1847 & & 0.9756 &  0.0245   \\
    Rectangle 4  & 0.750 & 1.150 & & -4.1531 & -0.0158 & & 1.0021 & -0.0022  \\
     \hline 
\end{tabular}
\end{threeparttable}
\end{table}


\section{Conclusions and Perspectives} \label{Conclusions}
In this work, we addressed the challenge of directly handling nonorthogonal spin orbitals in quantum algorithms, an issue of fundamental importance in quantum computing applications, due to the high computational costs traditionally associated with valence bond methods.

To this end, a modified Jordan-Wigner fermionic-to-spin mapping has been introduced. Implemented within \texttt{PennyLane}, this new encoding was shown to yield matrix element values between Slater determinants constructed from nonorthogonal spin orbitals in full agreement with conventional calculations based on the standard L\"{o}wdin rules. Furthermore, the application of the proposed mapping to the minimal model system H$_4$ produced chemically reasonable results, particularly in the coefficients and weights of the spin-coupled structures.

These preliminary findings pave the way to future development of quantum computing variants of valence bond methods. The fermionic-to-spin encoding presented here is expected to be central to such advancements. Equally important will be the design of a suitable \emph{ansatz} that preserves the characteristic chemical interpretability of traditional valence bond approaches. 

\section{Acknowledgements}
The Italian Research Center on High Performance Computing, Big Data and Quantum Computing (ICSC) is gratefully acknowledged for financial support of this work through the Innovation Grant Project ``Molecular Energy Landscapes by Quantum Computing – Benchmark Calculations". Ettore Canonici (Eni S.p.A.) is also thanked for helpful discussions.

\FloatBarrier




\bibliography{NonOrthoJW}
\bibliographystyle{elsarticle-num}





\end{document}